\documentclass[twocolumn,prl,floatfix]{revtex4}
\usepackage{amsmath}
\usepackage{bm}
\usepackage{euscript}
\usepackage{graphicx}
\graphicspath{{./Figures/}}

\begin{document}
\title{New experimental limits on non-Newtonian forces in the micrometer-range}
\author{A. O. Sushkov}
\email{alex.sushkov@yale.edu}
\affiliation{Yale University, Department of Physics, P.O. Box 208120, New Haven CT 06520-8120, USA}
\author{W. J. Kim}
\email{kimw@seattleu.edu}
\affiliation{Dept. of Physics, Seattle University, 901 12th Avenue, Seattle, WA 98122, USA}
\author{D. A. R. Dalvit}
\email{dalvit@lanl.gov}
\affiliation{Theoretical Division MS B213, Los Alamos National Laboratory, Los Alamos, NM 87545, USA}
\author{S. K. Lamoreaux}
\email{steve.lamoreaux@yale.edu}
\affiliation{Yale University, Department of Physics, P.O. Box 208120, New Haven, CT 06520-8120, USA}

\begin{abstract}
We report measurements of the short-range forces between two macroscopic gold-coated plates using a torsion pendulum. The force is measured for separations between 0.7~$\mu$m and 7~$\mu$m, and is well described by a combination of the Casimir force, including the finite-temperature correction, and an electrostatic force due to patch potentials on the plate surfaces. We use our data to place constraints on the Yukawa-type ``new'' forces predicted by theories with extra dimensions. We establish a new best bound for force ranges 0.4~$\mu$m to 4~$\mu$m, and, for forces mediated by gauge bosons propagating in $(4+n)$ dimensions and coupling to the baryon number, extract a $(4+n)$-dimensional Planck scale lower limit of $M_*>70$~TeV.
\end{abstract}

\date{\today}

\maketitle

It is remarkable that two of the greatest successes on 20th century physics, General Relativity and the Standard Model, appear to be fundamentally incompatible. Intense effort is devoted to searching for a framework that connects gravity to the rest of physics, and string theory, or M-theory, is a candidate. There is still a number of outstanding problems, two of the most serious ones are the gauge hierarchy problem and the cosmological constant problem. Theoretical approaches have included proposals incorporating $n$ extra spatial dimensions~\cite{Arkani-Hamed1999}, predicting deviations from Newtonian gravity at sub-millimeter length scales. The rationale is to bring down the Planck scale from $M_P=10^{19}$~GeV in 4 dimensions to the electroweak scale $M_*\approx 1$~TeV in $(4+n)$ dimensions, thereby addressing the gauge hierarchy problem. In addition, in this scenario, gauge bosons that propagate in the bulk of the $n$ extra dimensions but couple to the Standard Model baryon number, can mediate forces that are a factor of $\approx 10(M_*/M_N)^2\approx10^7$ stronger than gravity, here $M_N\approx 1$~GeV is nucleon mass. These forces have the Yukawa exponential form, with the range given by the Compton wavelength of the boson, proportional to the inverse of its mass, whose natural scale is $\approx M_*^2/M_P$, diluted, exactly like the gravitational interaction, by the bulk $(4+n)$-dimensional volume~\cite{Dimopoulos2003}.

A large amount of experimental work has been done to search for such forces in a wide range of distance scales~\cite{Adelberger2003}. The Yukawa potential due to new interactions is typically taken to modify the gravitational inverse-square law:
\begin{align}
\label{eq:V1}
V(r) = -G\frac{m_1m_2}{r}\left(1+\alpha e^{-r/\lambda}\right),
\end{align}
where $G$ is the gravitational constant, and the new interaction parameters are the strength $\alpha$ and the range $\lambda$. The strength $\alpha$ is constrained to be below unity for $\lambda>56$~$\mu$m~\cite{Kapner2007}, but at shorter ranges the experimental limits are not as stringent~\cite{Geraci2008,Masuda2009,Decca2005,Decca2007}. The measurements at short ranges are complicated by the presence of the Casimir force~\cite{Casimir1948,Onofrio2006}, as well as the electrostatic forces due to surface patch potentials~\cite{Speake2003,Kim2010}. See
\cite{Antoniadis2011} for a recent overview of tests of gravity at sub-millimeter ranges.

Recent measurements of the attractive force between two gold-coated flat and spherical plates for separations between 0.7~$\mu$m and 7~$\mu$m have improved our understanding of the Casimir and the electrostatic patch forces in this separation range, and detected the thermal Casimir force~\cite{Sushkov2011}. We now use these measurements to place limits on new interactions in the micron range.

Our apparatus, which has been more fully described in \cite{Sushkov2011} comprises a torsion pendulum suspended inside a vacuum chamber (pressure $5\times10^{-7}$~torr) by a tungsten wire of 25~$\mu$m diameter and 2.5~cm length. The force to be measured is between the two glass plates, each coated with a 700~\AA~(optically thick) layer of gold evaporated on top of a 100~\AA-thick layer of titanium. One is a flat plate mounted on one side of the pendulum,
the other is a spherical lens (radius of curvature $R=15.6$~cm, as measured with a Micromap TM-570 interferometric microscope at the Advanced Light Source Optical Metrology Laboratory~\cite{Yashchuk2006, Yashchuk2005}, and found to vary by less than 2\% over the surface of the lens), mounted on a Thorlabs T25 XYZ positioning stage, which, together with a piezoelectric transducer, is used to vary the plate separation $d$. The attractive force between the plates creates a torque on the pendulum body, which is counteracted by a pair of ``compensator'' electrodes on the opposite end of the pendulum. The voltage that has to be applied to the compensator electrodes to keep the pendulum stationary is proportional to the force between the Casimir plates, with the calibration coefficient extracted from the measurements of the electrostatic force between the plates.  Further details of the measurement technique can be found in Ref.~\cite{Sushkov2011}.

The total force between the plates can be written as:
\begin{align}
\label{eq:F1}
F = F_{\text{Casimir}}+F_{\text{electric}}+F_{\text{gravity}}+F_{\text{new}}.
\end{align}
The gravitational (Newtonian) force between the plates, $F_{\text{gravity}}$, is very nearly a constant ($\approx 20$~pN) in the studied range of separations, and is neglected in the analysis. $F_{\text{new}}$ is the hypothetical new force, arising from the Yukawa potential in Eq.~(\ref{eq:V1}). The Casimir force between the spherical lens and the planar plate is calculated in the proximity force approximation (valid for $d \ll R$) as
$F_{\text{Casimir}} =  2 \pi R E_{\text{Casimir}}$, where $E_{\text{Casimir}}$ is the Casimir interaction energy per unit area between two flat  parallel plates separated
by a distance $d$. The latter is computed using the Lifshitz formalism with temperature $T=300$~K, and the gold optical permittivity data~\cite{Palik1998}, extrapolated to zero frequency using the Drude model with parameters $\omega_p = 7.54$~eV, $\gamma = 0.051$~eV~\cite{Sushkov2011}.

The electrostatic force is given by the expression
\begin{align}
\label{eq:F2}
F_{\text{electric}}=\pi \epsilon_0 R \left[ \frac{(V-V_m)^2}{d} + \frac{V_{rms}^2}{d} \right],
\end{align}
where $\epsilon_0$ is the permittivity of free space,  $V$ is the computer-controlled bias voltage applied between the plates, and the ``minimizing potential'' offset $V_m$ is due to the contact potential difference of approximately 20~mV between the two plates, caused by the several solder contacts around the electrical loop connecting the two plates. Our measurements show that the minimizing potential $V_m(d)$ is nearly independent of separation in the 0.7~$\mu$m~$\leq d \leq$~7~$\mu$m range (average variation is 0.2~mV). $V_{rms}$ is a parameter characterizing the magnitude of the voltage fluctuations across the plates' surfaces, giving rise to a patch-potential electrostatic force given by the second term in brackets. Such voltage patches are always present even on chemically inert metal surfaces prepared in an ultra-clean environment~\cite{Robertson2006,Robertson2007}, and can be caused by spatial changes in surface crystalline structure, surface stresses, and adsorbed impurities or oxides.  The exact form of the electrostatic patch force is determined by the patch voltage size distribution spectrum on the plates
\cite{Speake2003}, and in particular by the relationship between three length scales: the typical patch size $\lambda$, the plate separation $d$, and the ``effective interaction length'' $r_{eff}=\sqrt{Rd}$. In the limit $d\ll \lambda \ll r_{eff}$ the patch force
is well described by $\pi \epsilon_0 R V_{rms}^2/d$~\cite{Kim2010}.


A further correction is needed to account for fluctuations in plate separation $d$~\cite{Lamoreaux2010}. The sources of these fluctuations are surface roughness of the plates, and pendulum fluctuations, caused, for example, by apparatus vibrations. In addition to radius of curvature measurements, surface roughness measurements were performed with the Micromap TM-570 interferometric microscope, yielding an rms roughness of $S_q\approx 10$~nm for the curved plate, and $S_q\approx 1$~nm for the flat plate. Vibration-caused fluctuations in $d$ were measured by connecting an inductor in parallel with the Casimir plates, and monitoring the resonance frequency of the resulting LC-circuit; rms fluctuations of $\lesssim 40$~nm were recorded. In addition, statistical error of $\pm 10$~nm in determination of $d$ contributes in quadrature to the fluctuations mentioned above. We take the total rms plate separation fluctuation of $\delta=(40\pm20)$~nm. From the Taylor expansion of the Casimir force about the mean plate separation, we deduce that a correction term $F_C''\delta^2/2$ has to be added to the theoretical force when comparing with experiment, the double prime denotes second-order derivative with respect to $d$. In addition, since the same correction exists for the electrostatic force, the plate separation $d$ extracted from the electrostatic calibration was corrected by a factor $1+(\delta/d)^2$, and the electrostatic patch force $V_{rms}^2/d$ was corrected by the same factor.

\begin{figure}[h!]
\includegraphics[width=\columnwidth]{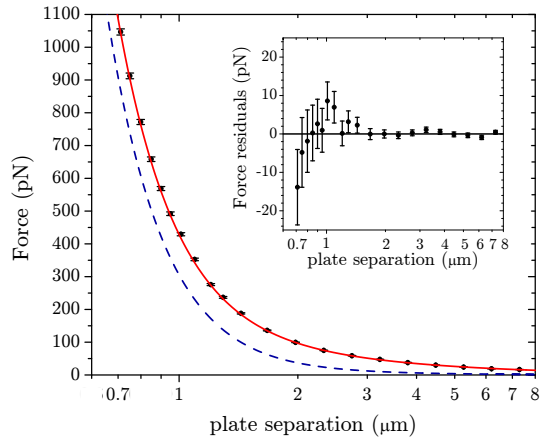}
\caption{The binned experimental short-range force between gold-coated plates. The error bars include contributions from statistical scatter, and uncertainties in the applied corrections, discussed in the text. The dashed blue line shows the theoretical Casimir force, calculated using the Lifshitz formalism at 300~K, with the Drude model permittivity extrapolation to zero frequency. The red line shows the force, including the electrostatic patch potential contribution, with two free fitting parameters, as described in the text. Inset: the force residuals, used to place constraints on the ``new'' short-range forces.}
\label{fig:TotalForce}
\end{figure}

The data are well described by the Drude model, using the distance correction derived from auxiliary measurements as described above (no free parameters), together with a least-squares fit for two parameters, which are $V_{rms}$ and an overall force offset.  Given that only two well-understood fitting parameters are needed to fully describe our data, which spans more that an order of magnitude in distance and more than two orders of magnitude in force, we are confident that, together with a $1/d$ patch potential force, the finite temperature Drude model provides the correct explanation of the Casimir force between Au surfaces. The reduced $\chi^2$ of the fit is 1.04. Therefore we can set bounds on additional forces that might be present, at a level of confidence based on the statistical fluctuations in the difference between the data and the corrected model.
The force data, grouped into distance bins and averaged, are shown in Figure~\ref{fig:TotalForce}, together with the best-fit line (red), and the Casimir force (dashed blue line). The difference between the red and blue curves is due to the patch potential $1/d$ force.  The fit residuals are shown in the inset.

\begin{figure}[h!]
\includegraphics[width=\columnwidth]{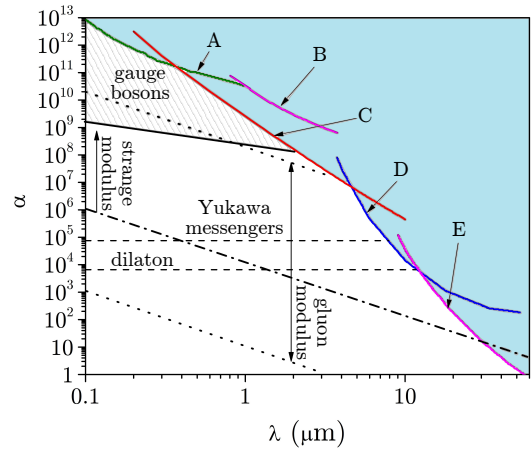}
\caption{Experimental upper limits on the Yukawa force strength $\alpha$, together with some theoretical predictions.
The area shaded in light blue is experimentally excluded. The curves labeled A-E correspond to results in Refs.~\cite{Decca2005}, \cite{Masuda2009}, present work, \cite{Geraci2008}, and \cite{Kapner2007}. The hatched area labeled ``gauge bosons'' is the parameter space for forces mediated by gauge bosons that propagate in $(4+n)$ dimensions and couple to the Standard Model baryon number.}
 \label{fig:YukawaLimits}
\end{figure}

According to Eq.~(\ref{eq:F1}), these residuals can be used to place a limit on the hypothetical ``new'' force $F_{\text{new}}$ between the plates. Integrating over the two gold- and titanium-coated plates gives the following approximate expression for the force:
\begin{align}
\label{eq:F3}
\begin{split}
F_{\text{new}} = & 4\pi^2 G R \alpha \lambda^3 e^{-d/\lambda}
[ \rho_{\text{Au}} + (\rho_{\text{Ti}}-\rho_{\text{Au}})e^{-d_{\text{Au}}/\lambda} \\
+ & (\rho_{\text{g}}-\rho_{\text{Ti}})e^{-(d_{\text{Au}}+d_{\text{Ti}})/\lambda} ]^2,
\end{split}
\end{align}
where $\rho_{\text{Au}}=19$~g/cm$^3$ is the gold density, $d_{\text{Au}}=700$~\AA~is the gold layer thickness, $\rho_{\text{Ti}}=4.5$~g/cm$^3$ is the Ti density, $d_{\text{Ti}}=100$~\AA~is the titanium layer thickness, and $\rho_{\text{g}}=2.6$~g/cm$^3$ is the substrate glass density. This expression is a good approximation to the exact form for the Yukawa force between the spherical lens and the flat plate provided $\lambda$, $d_{\text{Au}}$, and $d_{\text{Ti}}$ are much less than the curved plate's radius of curvature $R$, the flat plate's thickness, and both plates' diameters. These conditions are satisfied very well in our experiment (For an exact expression for the force $F_{\text{new}}$, not subject to these assumptions, see~\cite{exact}).  The obtained 95\%-confidence limits on the ``new'' interaction strength $\alpha$ at each interaction range $\lambda$ are shown in Figure~\ref{fig:YukawaLimits}. The figure also shows limits obtained by other experimental groups, as well as some theoretical expectations. Our experiment achieves up to a factor of 30 improvement in the limit on the interaction strength $\alpha$ for $0.4\;\mu\mathrm{m} < \lambda < 4\;\mu$m, compared to previous best limits~\cite{Masuda2009}.

Given the range of parameters $\alpha$, $\lambda$ that our experiment is most sensitive to, the most stringent limit we can place is on the $(4+n)$-dimensional Planck scale $M_*$ in presence of gauge bosons that propagate in $(4+n)$ dimensions and couple to the Standard Model baryon number (hatched region labeled ``gauge bosons'' in Fig.~\ref{fig:YukawaLimits}). Our data constrains the range of a hypothetical interaction mediated by such particles (i.e. their Compton wavelength) to be below 2~$\mu$m, which corresponds to the gauge particle mass of more than 0.5~eV. Assuming all the coupling parameters are on the order of unity, the natural scale for this mass is $M_*^2/M_P$, which means that the $(4+n)$-dimensional Planck scale is limited to $M_*>70$~TeV. This is more stringent than the astrophysical limits, based on the PSR J09052+0755 neutron star heating from Kaluza-Klein graviton decay, for the case of 3 or more extra dimensions~\cite{Hannestad2003}.

The authors thank Valery Yashchuk for performing the surface roughness measurements, and Roberto Onofrio and Serge Reynaud for discussions.
This work was supported by the DARPA/MTO's Casimir Effect Enhancement project under SPAWAR Contract No. N66001-09-1-2071.


\begin{thebibliography}{0}
\expandafter\ifx\csname natexlab\endcsname\relax\def\natexlab#1{#1}\fi
\expandafter\ifx\csname bibnamefont\endcsname\relax
  \def\bibnamefont#1{#1}\fi
\expandafter\ifx\csname bibfnamefont\endcsname\relax
  \def\bibfnamefont#1{#1}\fi
\expandafter\ifx\csname citenamefont\endcsname\relax
  \def\citenamefont#1{#1}\fi
\expandafter\ifx\csname url\endcsname\relax
  \def\url#1{\texttt{#1}}\fi
\expandafter\ifx\csname urlprefix\endcsname\relax\def\urlprefix{URL }\fi
\providecommand{\bibinfo}[2]{#2}
\providecommand{\eprint}[2][]{\url{#2}}

\end{thebibliography}


\begin{thebibliography}{17}

\expandafter\ifx\csname natexlab\endcsname\relax\def\natexlab#1{#1}\fi
\expandafter\ifx\csname bibnamefont\endcsname\relax
  \def\bibnamefont#1{#1}\fi
\expandafter\ifx\csname bibfnamefont\endcsname\relax
  \def\bibfnamefont#1{#1}\fi
\expandafter\ifx\csname citenamefont\endcsname\relax
  \def\citenamefont#1{#1}\fi
\expandafter\ifx\csname url\endcsname\relax
  \def\url#1{\texttt{#1}}\fi
\expandafter\ifx\csname urlprefix\endcsname\relax\def\urlprefix{URL }\fi
\providecommand{\bibinfo}[2]{#2}
\providecommand{\eprint}[2][]{\url{#2}}

\bibitem[{\citenamefont{{Arkani-Hamed}
  et~al.}(1999)\citenamefont{{Arkani-Hamed}, Dimopoulos, and
  Dvali}}]{Arkani-Hamed1999}
\bibinfo{author}{\bibfnamefont{N.}~\bibnamefont{{Arkani-Hamed}}},
  \bibinfo{author}{\bibfnamefont{S.}~\bibnamefont{Dimopoulos}},
  \bibnamefont{and} \bibinfo{author}{\bibfnamefont{G.}~\bibnamefont{Dvali}},
  \bibinfo{journal}{Phys. Rev.  D} \textbf{\bibinfo{volume}{59}},
  \bibinfo{pages}{086004} (\bibinfo{year}{1999}).

\bibitem[{\citenamefont{Dimopoulos and Geraci}(2003)}]{Dimopoulos2003}
\bibinfo{author}{\bibfnamefont{S.}~\bibnamefont{Dimopoulos}} \bibnamefont{and}
  \bibinfo{author}{\bibfnamefont{A.~A.} \bibnamefont{Geraci}},
  \bibinfo{journal}{Phys. Rev.  D} \textbf{\bibinfo{volume}{68}},
  \bibinfo{pages}{124021} (\bibinfo{year}{2003}).

\bibitem[{\citenamefont{Adelberger et~al.}(2003)\citenamefont{Adelberger,
  Heckel, and Nelson}}]{Adelberger2003}
\bibinfo{author}{\bibfnamefont{E.}~\bibnamefont{Adelberger}},
  \bibinfo{author}{\bibfnamefont{B.}~\bibnamefont{Heckel}}, \bibnamefont{and}
  \bibinfo{author}{\bibfnamefont{A.}~\bibnamefont{Nelson}},
  \bibinfo{journal}{Annual Review of Nuclear and Particle Science}
  \textbf{\bibinfo{volume}{53}}, \bibinfo{pages}{77} (\bibinfo{year}{2003}).

\bibitem[{\citenamefont{Kapner et~al.}(2007)\citenamefont{Kapner, Cook,
  Adelberger, Gundlach, Heckel, Hoyle, and Swanson}}]{Kapner2007}
\bibinfo{author}{\bibfnamefont{D.~J.} \bibnamefont{Kapner}},
  \bibinfo{author}{\bibfnamefont{T.~S.} \bibnamefont{Cook}},
  \bibinfo{author}{\bibfnamefont{E.~G.} \bibnamefont{Adelberger}},
  \bibinfo{author}{\bibfnamefont{J.~H.} \bibnamefont{Gundlach}},
  \bibinfo{author}{\bibfnamefont{B.~R.} \bibnamefont{Heckel}},
  \bibinfo{author}{\bibfnamefont{C.~D.} \bibnamefont{Hoyle}}, \bibnamefont{and}
  \bibinfo{author}{\bibfnamefont{H.~E.} \bibnamefont{Swanson}},
  \bibinfo{journal}{Phys. Rev. Lett.} \textbf{\bibinfo{volume}{98}},
  \bibinfo{pages}{021101} (\bibinfo{year}{2007}).

\bibitem[{\citenamefont{Geraci et~al.}(2008)\citenamefont{Geraci, Smullin,
  Weld, Chiaverini, and Kapitulnik}}]{Geraci2008}
\bibinfo{author}{\bibfnamefont{A.~A.} \bibnamefont{Geraci}},
  \bibinfo{author}{\bibfnamefont{S.~J.} \bibnamefont{Smullin}},
  \bibinfo{author}{\bibfnamefont{D.~M.} \bibnamefont{Weld}},
  \bibinfo{author}{\bibfnamefont{J.}~\bibnamefont{Chiaverini}},
  \bibnamefont{and}
  \bibinfo{author}{\bibfnamefont{A.}~\bibnamefont{Kapitulnik}},
  \bibinfo{journal}{Phys. Rev.  D} \textbf{\bibinfo{volume}{78}},
  \bibinfo{pages}{022002} (\bibinfo{year}{2008}).

\bibitem[{\citenamefont{Masuda and Sasaki}(2009)}]{Masuda2009}
\bibinfo{author}{\bibfnamefont{M.}~\bibnamefont{Masuda}} \bibnamefont{and}
  \bibinfo{author}{\bibfnamefont{M.}~\bibnamefont{Sasaki}},
  \bibinfo{journal}{Phys. Rev. Lett.} \textbf{\bibinfo{volume}{102}},
  \bibinfo{pages}{171101} (\bibinfo{year}{2009}).

\bibitem[{\citenamefont{Decca et~al.}(2005)\citenamefont{Decca, L\'opez, Chan,
  Fischbach, Krause, and Jamell}}]{Decca2005}
\bibinfo{author}{\bibfnamefont{R.~S.} \bibnamefont{Decca}},
  \bibinfo{author}{\bibfnamefont{D.}~\bibnamefont{L\'opez}},
  \bibinfo{author}{\bibfnamefont{H.~B.} \bibnamefont{Chan}},
  \bibinfo{author}{\bibfnamefont{E.}~\bibnamefont{Fischbach}},
  \bibinfo{author}{\bibfnamefont{D.~E.} \bibnamefont{Krause}},
  \bibnamefont{and} \bibinfo{author}{\bibfnamefont{C.~R.}
  \bibnamefont{Jamell}}, \bibinfo{journal}{Phys. Rev. Lett.}
  \textbf{\bibinfo{volume}{94}}, \bibinfo{pages}{240401}
  (\bibinfo{year}{2005}).

  \bibitem{Decca2007}
  R.S. Decca, D. L\'opez, E. Fischbach, G.L. Klimchitskaya, D.E. Krause, and V.M. Mostepanenko,
  Phys. Rev. D {\bf 75}, 077101 (2007).

\bibitem[{\citenamefont{Casimir}(1948)}]{Casimir1948}
\bibinfo{author}{\bibfnamefont{H.~B.~G.} \bibnamefont{Casimir}},
  \bibinfo{journal}{Proc. Kon. Nederland. Akad. Wetensch.}
  \textbf{\bibinfo{volume}{51}}, \bibinfo{pages}{793} (\bibinfo{year}{1948}).

\bibitem{Onofrio2006}
R. Onofrio, New. J. Phys. {\bf 8}, 237 (2006).

\bibitem{Speake2003}
C.C. Speake and C. Trenkel, Phys. Rev. Lett. {\bf 90}, 160403 (2003).

\bibitem[{\citenamefont{Kim et~al.}(2010)\citenamefont{Kim, Sushkov, Dalvit,
  and Lamoreaux}}]{Kim2010}
\bibinfo{author}{\bibfnamefont{W.~J.} \bibnamefont{Kim}},
  \bibinfo{author}{\bibfnamefont{A.~O.} \bibnamefont{Sushkov}},
  \bibinfo{author}{\bibfnamefont{D.~A.~R.} \bibnamefont{Dalvit}},
  \bibnamefont{and} \bibinfo{author}{\bibfnamefont{S.~K.}
  \bibnamefont{Lamoreaux}}, \bibinfo{journal}{Phys. Rev. A}
  \textbf{\bibinfo{volume}{81}}, \bibinfo{pages}{022505}
  (\bibinfo{year}{2010}).

\bibitem{Antoniadis2011}
I. Antoniadis, S. Baessler, M. Buchner, V.V. Fedorov, S. Hoedl, V.V. Nesvizhevsky, G.
Pignol, K.V. Protasov, S. Reynaud and Yu. Sobolev, Compt. Rend. Acad. Sci. (to appear, 2011).

\bibitem[{\citenamefont{Sushkov et~al.}(2011)\citenamefont{Sushkov, Kim,
  Dalvit, and Lamoreaux}}]{Sushkov2011}
\bibinfo{author}{\bibfnamefont{A.~O.} \bibnamefont{Sushkov}},
  \bibinfo{author}{\bibfnamefont{W.~J.} \bibnamefont{Kim}},
  \bibinfo{author}{\bibfnamefont{D.~A.~R.} \bibnamefont{Dalvit}},
  \bibnamefont{and} \bibinfo{author}{\bibfnamefont{S.~K.}
  \bibnamefont{Lamoreaux}}, \bibinfo{journal}{Nat. Phys.}
  \textbf{\bibinfo{volume}{7}}, \bibinfo{pages}{230} (\bibinfo{year}{2011}).

\bibitem[{\citenamefont{Yashchuk et~al.}(2006)\citenamefont{Yashchuk,
  Gullikson, Howells, Irick, {MacDowell}, {McKinney}, Salmassi, Warwick, Metz,
  and Tonnessen}}]{Yashchuk2006}
\bibinfo{author}{\bibfnamefont{V.~V.} \bibnamefont{Yashchuk}},
  \bibinfo{author}{\bibfnamefont{E.~M.} \bibnamefont{Gullikson}},
  \bibinfo{author}{\bibfnamefont{M.~R.} \bibnamefont{Howells}},
  \bibinfo{author}{\bibfnamefont{S.~C.} \bibnamefont{Irick}},
  \bibinfo{author}{\bibfnamefont{A.~A.} \bibnamefont{{MacDowell}}},
  \bibinfo{author}{\bibfnamefont{W.~R.} \bibnamefont{{McKinney}}},
  \bibinfo{author}{\bibfnamefont{F.}~\bibnamefont{Salmassi}},
  \bibinfo{author}{\bibfnamefont{T.}~\bibnamefont{Warwick}},
  \bibinfo{author}{\bibfnamefont{J.~P.} \bibnamefont{Metz}}, \bibnamefont{and}
  \bibinfo{author}{\bibfnamefont{T.~W.} \bibnamefont{Tonnessen}},
  \bibinfo{journal}{Applied Optics} \textbf{\bibinfo{volume}{45}},
  \bibinfo{pages}{4833} (\bibinfo{year}{2006}).

\bibitem[{\citenamefont{Yashchuk et~al.}(2005)\citenamefont{Yashchuk, Franck,
  Irick, Howells, {MacDowell}, and {McKinney}}}]{Yashchuk2005}
\bibinfo{author}{\bibfnamefont{V.~V.} \bibnamefont{Yashchuk}},
  \bibinfo{author}{\bibfnamefont{A.~D.} \bibnamefont{Franck}},
  \bibinfo{author}{\bibfnamefont{S.~C.} \bibnamefont{Irick}},
  \bibinfo{author}{\bibfnamefont{M.~R.} \bibnamefont{Howells}},
  \bibinfo{author}{\bibfnamefont{A.~A.} \bibnamefont{{MacDowell}}},
  \bibnamefont{and} \bibinfo{author}{\bibfnamefont{W.~R.}
  \bibnamefont{{McKinney}}}, in \emph{\bibinfo{booktitle}{Nano- and
  {Micro-Metrology}}}, edited by
  \bibinfo{editor}{\bibfnamefont{H.}~\bibnamefont{Ottevaere}},
  \bibinfo{editor}{\bibfnamefont{P.}~\bibnamefont{{DeWolf}}}, \bibnamefont{and}
  \bibinfo{editor}{\bibfnamefont{D.~S.} \bibnamefont{Wiersma}}
  (\bibinfo{publisher}{{SPIE}}, \bibinfo{address}{Munich, Germany},
  \bibinfo{year}{2005}), vol. \bibinfo{volume}{5858}, pp.
  \bibinfo{pages}{58580A--12}.

\bibitem[{\citenamefont{Palik}(1998)}]{Palik1998}
\bibinfo{editor}{\bibfnamefont{E.~D.} \bibnamefont{Palik}}, ed.,
  \emph{\bibinfo{title}{Handbook of Optical Constants of Solids}}
  (\bibinfo{publisher}{Elsevier}, \bibinfo{year}{1998}).

\bibitem[{\citenamefont{Robertson et~al.}(2006)\citenamefont{Robertson,
  Blackwood, Buchman, Byer, Camp, Gill, Hanson, Williams, and
  Zhou}}]{Robertson2006}
\bibinfo{author}{\bibfnamefont{N.~A.} \bibnamefont{Robertson}},
  \bibinfo{author}{\bibfnamefont{J.~R.} \bibnamefont{Blackwood}},
  \bibinfo{author}{\bibfnamefont{S.}~\bibnamefont{Buchman}},
  \bibinfo{author}{\bibfnamefont{R.~L.} \bibnamefont{Byer}},
  \bibinfo{author}{\bibfnamefont{J.}~\bibnamefont{Camp}},
  \bibinfo{author}{\bibfnamefont{D.}~\bibnamefont{Gill}},
  \bibinfo{author}{\bibfnamefont{J.}~\bibnamefont{Hanson}},
  \bibinfo{author}{\bibfnamefont{S.}~\bibnamefont{Williams}}, \bibnamefont{and}
  \bibinfo{author}{\bibfnamefont{P.}~\bibnamefont{Zhou}},
  \bibinfo{journal}{Classical and Quantum Gravity}
  \textbf{\bibinfo{volume}{23}}, \bibinfo{pages}{2665} (\bibinfo{year}{2006}).

\bibitem[{\citenamefont{Robertson}(2007)}]{Robertson2007}
\bibinfo{author}{\bibfnamefont{N.~A.} \bibnamefont{Robertson}},
  \bibinfo{journal}{Report {LIGO-G070481-00-R} (available at
  {http://www.ligo.caltech.edu/docs/G/G070481-00.pdf)}}
  (\bibinfo{year}{2007}).

\bibitem[{\citenamefont{Lamoreaux}(2010)}]{Lamoreaux2010}
\bibinfo{author}{\bibfnamefont{S.~K.} \bibnamefont{Lamoreaux}},
  \bibinfo{journal}{Phys. Rev. A} \textbf{\bibinfo{volume}{82}},
  \bibinfo{pages}{024102} (\bibinfo{year}{2010}).

\bibitem{exact}
M. Bordag, B. Geyer, G.L. Klimchitskaya, and V.M. Mostepanenko, Phys. Rev. D {\bf 62}, 011701(R) (2000);
D.A.R. Dalvit and R. Onofrio, Phys. Rev. D {\bf 80}, 064025 (2009);
E. Fishbach, G.L. Klimchitskaya, D.E. Krause, and V.M. Mostepanenko,
Eur. Phys. J. C {\bf 68}, 223 (2010).

\bibitem[{\citenamefont{Hannestad and Raffelt}(2003)}]{Hannestad2003}
\bibinfo{author}{\bibfnamefont{S.}~\bibnamefont{Hannestad}} \bibnamefont{and}
  \bibinfo{author}{\bibfnamefont{G.~G.} \bibnamefont{Raffelt}},
  \bibinfo{journal}{Phys. Rev. D} \textbf{\bibinfo{volume}{67}},
  \bibinfo{pages}{125008} (\bibinfo{year}{2003}).

\end{thebibliography}

\end{document}